\documentclass[aps,onecolumn,11pt,floatfix,altaffilletter,tightenlines,showpacs,showkeys,notitlepage,nofootinbib]{revtex4-1}
\pdfoutput=1
\usepackage[colorlinks=true,citecolor=blue,linkcolor=blue]{hyperref}
\usepackage[normalem]{ulem}
\usepackage{amsmath,amssymb}
\usepackage{enumitem}
\usepackage{relsize}
\usepackage{mathtools}
\usepackage{epsfig}
\usepackage{graphicx}               
\usepackage{url}
\usepackage{color}
\usepackage{multirow}
\usepackage{placeins}
\usepackage[dvipsnames]{xcolor}
\usepackage{epstopdf}
\usepackage{nicefrac}
\usepackage{siunitx}
\usepackage{csquotes}

\usepackage[capitalize]{cleveref}
\usepackage{lipsum}
 \usepackage{gensymb}
 
\allowdisplaybreaks

\bibpunct{[}{]}{,}{n}{}{,}

\newcommand{\dd}{\mbox{d}}
\newcommand{\E}{\mathrm{e}}
\newcommand{\I}{\mathrm{i}}

\newcommand{\ie}{\emph{i.e.}~}
\newcommand{\eg}{\emph{e.g.}~}

\newcommand{\transpose}{\intercal}

\newcommand{\tinytext}[1]{\text{\tiny{#1}}}
\newcommand{\fineq}[1]{\;#1}

\newcommand{\MPl}{M_\tinytext{Pl}}
\newcommand{\hubble}{\mathcal{H}}
\newcommand{\fieldVec}{\vec{\Phi}}

\newcommand{\oN}[1]{\text{O}(#1)}


\begin{document}

\title{Strong Supercooling as a Consequence of Renormalization Group Consistency}

\author{Vedran Brdar}   \email{vbrdar@mpi-hd.mpg.de}
\author{Alexander J.~Helmboldt} \email{alexander.helmboldt@mpi-hd.mpg.de}
\author{Manfred Lindner} \email{lindner@mpi-hd.mpg.de}
\affiliation{Max-Planck-Institut f\"ur Kernphysik,
       69117~Heidelberg, Germany}

\begin{abstract}
\noindent
Classically scale-invariant models are attractive not only because they may
offer a solution to the long-standing gauge hierarchy problem, but also
due to their role in facilitating strongly supercooled cosmic phase
transitions.
In this paper, we investigate the interplay between these two aspects.
We do so in the context of the electroweak phase transition (EWPT) in
the minimal scale-invariant theory.
We find that the amount of supercooling generally decreases for
increasing scalar couplings.
However, the stabilization of the electroweak scale against the Planck
scale requires the absence of Landau poles in the respective energy
range.
Scalar couplings at the TeV scale can therefore not become larger than
 $\mathcal{O}(10^{-1})$.
As a consequence, all fully consistent parameter points predict the EWPT
not to complete before the QCD transition, which then eventually
triggers the generation of the electroweak scale.
We also discuss the potential of the model to give rise to an observable
gravitational wave signature, as well as the possibility to accommodate
a dark matter candidate.
\end{abstract}

\maketitle

\section{Motivation}
\label{sec:intro}
\noindent
The absence of new physics discoveries at the LHC leaves the gauge hierarchy problem unsolved, so that the origin of the smallness of the Higgs mass remains unknown.
While solutions relying on low-scale supersymmetry are still possible, it certainly becomes increasingly relevant to explore viable alternative scenarios.
One such approach is built around models based on classical scale invariance
\cite{Bardeen:1995kv,Hempfling:1996ht,Meissner:2006zh,Meissner:2007xv,Meissner:2009gs,
Foot:2007as,Foot:2007ay},
in which the Lagrangian contains exclusively dimensionless parameters and where all mass scales arise by dimensional transmutation.
Apart from their connection to the hierarchy problem, classically scale-invariant theories recently received some attention because of the role they could play in facilitating strongly supercooled phase transitions with various interesting implications for the physics of the early Universe and the detection of gravitational waves \cite{Marzola:2017jzl,Iso:2017uuu,Hambye:2018qjv,Baldes:2018emh,Prokopec:2018tnq,Brdar:2018num,Marzo:2018nov,Mohamadnejad:2019vzg}.
In the present paper we will investigate the interplay between the aforementioned two crucial features of scale-invariant models more closely.

To set the stage, let us briefly review the most important aspects of scale-invariant theories, in particular their relation to the hierarchy problem and supercooled phase transitions.
At the core of the considered class of models is the dynamical breaking of scale invariance, which is essential to explaining the existence of massive particles.
It can proceed in two complementary ways: either in the perturbative or in the strong coupling regime.
Here, we will focus on the former approach, where scale invariance is radiatively broken by renormalization group (RG) effects \`{a} la Coleman-Weinberg \cite{CW,Weinberg:1973am}.
It is well-known that this type of symmetry breaking works in the minimal Standard Model (SM), but is phenomenologically excluded due to the large observed Higgs and top quark masses.
Consequently, viable theories must extend the SM by additional scalar degrees of freedom.
These scalar fields can then dynamically acquire a finite vacuum expectation value (VEV) thus spontaneously breaking classical scale invariance.
The created scale subsequently needs to be communicated to the SM sector, which is usually realized via the portal term between the SM Higgs doublet and the new scalar(s), see \eg Ref.~\cite{Meissner:2006zh}.
Alternatively, the electroweak scale can be radiatively generated from quantum fluctuations of heavy right-handed neutrinos \cite{Brivio:2017dfq,Brdar:2018vjq,Brivio:2018rzm,Brdar:2019iem,Brivio:2019hrj}, which simultaneously produce light active neutrino masses via the type-I seesaw mechanism \cite{Goran,Yanagida:1979as,GellMann:1980vs,Minkowski}.
Interestingly, both methods of scale transmission are anticipated to be testable by gravitational wave detectors \cite{Marzola:2017jzl,Brdar:2018num}.
In this work, we will concentrate on the minimal portal model \cite{Helmboldt:2016mpi,AlexanderNunneley:2010nw}, which can additionally be probed at colliders.
The study of radiative symmetry breaking in this model effectively boils down to the minimization of a multi-scalar effective potential.
For this purpose an elegant, yet approximative method proposed by Gildener and Weinberg is often employed \cite{GW}, which relies on the existence of an exact flat direction in the model's tree-level potential.
In order to make sure that our findings are not an artifact of such a strong assumption, we implemented and used a numerical method that does not depend on similar approximations.

Regardless of the concrete implementation of classical scale invariance and the numerical treatment of its breaking, its connection to resolving the hierarchy problem always relies on the following property:
In the absence of explicit mass scales in the underlying theory, the anomalous breaking of scale invariance by quantum effects only introduces a logarithmic sensitivity of infrared to ultraviolet physics \cite{Bardeen:1995kv}, where the latter is usually assumed to be connected to gravity.
In order for the above statement to remain true, there cannot be any physical thresholds between the scale of radiative symmetry breaking and the Planck scale \cite{Meissner:2007xv}.
In particular, this implies that the electroweak scale can only be stabilized against the Planck scale in a classically scale-invariant model, if the RG flow of the theory's couplings across said energy range is free of Landau poles \cite{Meissner:2006zh,Meissner:2007xv}.

As mentioned earlier, classical scale-invariance is also interesting because of its connection to strongly supercooled cosmic phase transitions (PT).
To appreciate this feature, let us mention that the amount of supercooling in a successful first-order PT \textendash\ and thus its strength as well as that of the associated gravitational wave signal \cite{Witten1984,Hogan1984,Hogan1986b,Turner1990a,Kamionkowski1993} \textendash\ is limited in conventional models built around polynomial scalar potentials \cite{Ellis:2018mja}.
Scale-invariant theories, on the other hand, are based on a nearly-conformal effective potential and can thus circumvent the aforementioned upper limits \cite{Konstandin:2011dr}.
Depending on the concrete scenario, supercooling can even become so strong that the scale-symmetry-breaking PT does not complete until the Universe cools down to temperatures of the order of the QCD scale \cite{Iso:2017uuu,Hambye:2018qjv,Baldes:2018emh}.
In such cases, the chiral PT of QCD is anticipated to proceed anyway, which will eventually induce electroweak symmetry breaking \cite{Iso:2017uuu}.

The paper is organized as follows.
In \cref{subsec:model}, we present the basics of the minimal phenomenologically viable scale-invariant model at zero temperature.
\cref{subsec:magic} then outlines our procedure for studying radiative scale symmetry breaking and the subsequent generation of the electroweak scale.
In \cref{subsec:finiteT} we incorporate finite-temperature effects into the minimal model by calculating the daisy-corrected thermal one-loop effective potential.
We furthermore give a brief general discussion on the scale-symmetry-breaking phase transition.
In \cref{subsec:main}, we investigate how the realization of strong supercooling is influenced by requiring RG consistency.
Potential gravitational wave signatures are discussed in \cref{subsec:DM-GW}, where we also comment on whether a viable dark matter (DM) candidate can be accommodated.
We finally conclude in \cref{sec:summary}.

\section{Minimal Viable Classically Scale-Invariant Realization}
\label{sec:2}

\subsection{The model at zero temperature}
\label{subsec:model}
\noindent
The minimal classically scale-invariant extension of the Standard Model (SM) which is consistent with current phenomenological observations and which can avoid Landau poles below the Planck scale, features two real scalar gauge singlets \cite{Helmboldt:2016mpi}.
The model's tree-level potential is given by
\begin{align}
	V_\text{tree}(H,S,R) &={} \lambda (H^\dagger H)^2 + \tfrac{1}{4}\lambda_s S^4 + \tfrac{1}{4} \lambda_r R^4 
	 + \tfrac{1}{2}\lambda_{\phi s} (H^\dagger H) S^2 + \tfrac{1}{2}\lambda_{\phi r} (H^\dagger H) R^2 + \tfrac{1}{4} \lambda_{sr} S^2 R^2 
	\fineq{,}
	\label{eq:Vtree}
\end{align}
where $S$ and $R$ denote these novel scalar degrees of freedom. In order to simplify the potential we assumed the existence of a $\mathbb{Z}_2$ symmetry under which $R$ transforms non-trivially\footnote{Such a choice can be motivated by dark matter (DM) stability. Indeed, we will comment on the possibility of explaining DM with the considered model in \cref{subsec:DM-GW}.},
such that the terms odd in $R$ are absent.
In \cref{eq:Vtree}, the SM Higgs doublet $H$ can be parametrized in terms of real fields, namely
\begin{align}
	H = \tfrac{1}{\sqrt{2}}
	\begin{pmatrix}
		\chi_1 + \I \chi_2 \\
		\phi_c + \phi + \I \chi_3 \\
	\end{pmatrix}
	\fineq{,}
\end{align}
where $\phi$ denotes the neutral CP-even Higgs field, while the $\chi_i$ with \mbox{$i=1,2,3$} represent the Goldstone bosons.
The classical field $\phi_c$ converges in vacuum toward \mbox{$v_\phi=\SI{246}{GeV}$}.
Similarly, we parametrize \mbox{$S=s_c+s$}, with a fluctuation field $s$ and a background field $s_c$ that approaches a finite value $v_s$ in the vacuum, thus spontaneously breaking classical scale invariance.
Furthermore, in accordance with the discussion in Ref.~\cite{Helmboldt:2016mpi}, we will only be interested in parameter points where the $R$ singlet does not acquire a finite vacuum expectation value (VEV), \ie where the $\mathbb{Z}_2$ symmetry introduced above is also a symmetry of the true vacuum.
Correspondingly, we write $R=r$ in what follows.
Based on the described symmetry breaking pattern, one can now identify the terms in the potential \eqref{eq:Vtree} from which the masses of the scalar particles arise.
We write
\begin{align}
V_\text{tree} \supseteq \tfrac{1}{2} \begin{pmatrix} \phi & s \end{pmatrix} m^2 \begin{pmatrix} \phi \\ s \end{pmatrix}
	+ \tfrac{1}{2} m_r^2 r^2 + \tfrac{1}{2} m_\chi^2 \chi_i^2 \fineq{,}
\end{align}
with the CP-even scalars' mass-squared matrix
\begin{align}
	m^2 \equiv m^2(\phi_c, s_c) =
	\begin{pmatrix}
		3\lambda \phi_c^2 + \tfrac{1}{2}\lambda_{\phi s} s_c^2 & \lambda_{\phi s} \phi_c s_c \\
		\lambda_{\phi s} \phi_c s_c & 3\lambda_s s_c^2 + \tfrac{1}{2}\lambda_{\phi s} \phi_c^2 \\
	\end{pmatrix}\equiv
	\begin{pmatrix}
	\mathcal{A} & \mathcal{B} \\
	\mathcal{B} & \mathcal{C}
	\end{pmatrix}
	\fineq{,}
\label{eq:mass-higgs}
\end{align}
where the last equality is introduced for later convenience.
The field-dependent tree-level masses for $r$ and the Goldstone bosons read
\begin{align}
	m^2_r(\phi_c, s_c) = \tfrac{1}{2}(\lambda_{\phi r} \phi_c^2 + \lambda_{sr} s_c^2)
	\qquad\text{and}\qquad
	m^2_\chi(\phi_c, s_c) = \lambda \phi_c^2 + \tfrac{1}{2}\lambda_{\phi s} s_c^2 \fineq{.}
	\label{eq:R-Goldstone}
\end{align}
In order to obtain the masses of the remaining CP-even scalars, we diagonalize \cref{eq:mass-higgs} and obtain
\begin{align}
	\begin{split}
	m^2_\pm(\phi_c,s_c) ={}& \frac{1}{4} \Bigl[ (6\lambda + \lambda_{\phi s}) \phi_c^2 + (6\lambda_s + \lambda_{\phi s}) s_c^2 \\
	& \pm \sqrt{[(6\lambda-\lambda_{\phi s})\phi_c^2 - (6\lambda_s-\lambda_{\phi s})s_c^2]^2 + 16\lambda_{\phi s}^2 \phi_c^2 s_c^2} \Bigr] \fineq{,}
	\end{split}
	\label{eq:pm} 
\end{align}
where the $+$ $(-)$ subscript denotes the larger (smaller) eigenvalue.

For the phase transition analysis we include the full set of one-loop corrections to the tree-level scalar potential.
While the temperature-dependent terms will be introduced in \cref{subsec:finiteT}, here we define the usual Coleman-Weinberg potential \cite{CW} employing the $\overline{\text{MS}}$ renormalization scheme and Landau gauge
\begin{align}
	V_\tinytext{CW}(\phi_c,s_c) = \frac{1}{64\pi^2} \sum_i n_i \, m_i^4(\phi_c,s_c) \left( \log\frac{m_i^2(\phi_c,s_c)}{\bar{\mu}^2} - c_i \right) \fineq{,}
	\label{eq:CW}
\end{align}
so that \mbox{$c_i=3/2$} for fermions and scalars, whereas \mbox{$c_i=5/6$} for gauge bosons.
Additionally, we fix the renormalization scale $\bar{\mu}$ such that \mbox{$\bar{\mu}^2=v_s^2+v_\phi^2$}.
The number of real degrees of freedom of particle species $i$ (including an additional minus sign for fermionic fields) is denoted as $n_i$.
The sum in \cref{eq:CW} is taken over all relevant fields, $i\in\{+,-,r,\chi,t,W,Z\}$, thereby also consistently including the leading fermionic (top quark) and gauge boson ($W$ and $Z$) SM contributions associated with the following field-dependent tree-level masses
\begin{align}
	m_t^2(\phi_c) & = \tfrac{1}{2} y_t^2 \,\phi_c^2 \fineq{,} &
	m_W^2(\phi_c) & = \tfrac{1}{4} g^2 \,\phi_c^2 \fineq{,} &
	m_Z^2(\phi_c) & = \tfrac{1}{4}(g^2 + {g^\prime}^2) \,\phi_c^2 \fineq{.}  
	\label{eq:cSM2S:nonscalar_masses}
\end{align}
Here, $g$ and $g'$ are $SU(2)_L$ and $U(1)_Y$ gauge coupling constants and $y_t$ is the top quark Yukawa coupling. Note that top quark, $W$, $Z$ and would-be Goldstone boson contributions enter in \cref{eq:CW} with prefactors of  $n_i=-12, 6, 3,\text{ and }3$, respectively, while \mbox{$n_i=1$} otherwise.

\subsection{Finding Consistent Parameter Sets}
\label{subsec:magic}
\noindent
In \cref{eq:Vtree} we introduced six quartic couplings, among which only $\lambda_r$ does not enter in the expressions for scalar tree-level masses. That coupling will be set to zero at the renormalization point $\bar{\mu}$ throughout the analysis.
In what follows, we will briefly describe our strategy for determining the remaining couplings.
The input parameters are $v_s$, $v_\phi$ and the (tree-level) mixing angle that diagonalizes the mass-squared matrix of $\phi$ and $s$, denoted $\theta_p$.
Given that \mbox{$v_\phi=\SI{246}{GeV}$}, the magnitude of $v_s$ is conveniently regulated by 
another dimensionless parameter $\theta_m$ which is defined as \mbox{$\theta_m=\arctan(v_\phi/v_s)$}. The following three conditions determine $\lambda$, $\lambda_{\phi s}$ and $\lambda_s$ unambiguously:
\begin{enumerate}
	\item[$(i)$] Working in Landau gauge, the masses of the would-be Goldstone bosons evaluated at the vacuum need to (approximately) vanish, which implies \mbox{$\lambda_{\phi s}\simeq-2 \lambda \,\tan^2\theta_m$} according to \cref{eq:R-Goldstone} and the definition of $\theta_m$.
	\item[$(ii)$] The neutral scalar mass-squared matrix of \cref{eq:mass-higgs} is diagonalized if the mixing angle $\theta_p$ is defined as $\theta_p = \frac{1}{2} \arctan \frac{2\,\mathcal{B}}{\mathcal{A}-\mathcal{C}}$, where $\mathcal{A}$, $\mathcal{B}$, and $\mathcal{C}$ were defined in \cref{eq:mass-higgs}.
	\item[$(iii)$] Either $m_+^2$ or $m_-^2$ evaluated at the vacuum needs to be approximately \mbox{$m_\text{Higgs}^2=(\SI{125}{GeV})^2$}. To be more precise, one of the CP-even scalars needs to have the mass and the couplings of the observed Higgs boson.
\end{enumerate}
By using $(i)$ and $(ii)$, one can derive the expression \mbox{$\mathcal{C}=\mathcal{A}\,(1+2 \tan \theta_m/\tan 2 \theta_p)$}, which nicely illustrates the relation between $(ii)$ and $(iii)$. Namely, in the phenomenological limit of small mixing, $\mathcal{A}$ is approximately equal to the mass of SM Higgs boson, whereas $\mathcal{C}$ corresponds to the mass of another CP-even scalar.
Given that $\theta_m$ is in the range $(0,\pi/2)$, it is the sign of $\theta_p$ which determines whether the SM-like Higgs is the lighter or heavier CP-even boson.
From \mbox{$\theta_p>0$}, it follows that \mbox{$m_\text{Higgs}^2=m_-^2$}, whereas \mbox{$\theta_p<0$} implies \mbox{$m_\text{Higgs}^2=m_+^2$}.

After determining $\lambda$, $\lambda_{\phi s}$ and $\lambda_s$, we still need to fix the values of the portal couplings of the $r$ field, namely $\lambda_{sr}$ and $\lambda_{\phi r}$.
We infer these values from the two stationarity conditions 
\begin{align}
	\left. \frac{\partial V(\phi,v_s)}{\partial \phi} \right|_{\phi=v_\phi}=0
	\qquad\text{and}\qquad
	\left. \frac{\partial V(v_\phi,s)}{\partial s} \right|_{s=v_s} = 0 \fineq{,}
	\label{eq:first_der}
\end{align}
where $V=V_\text{tree}+V_\tinytext{CW}$, see \cref{subsec:model} for the definitions.
Furthermore, it is crucial to check that the Hessian matrix of $V$ is positive definite.
In that case the parameters obtained from \cref{eq:first_der} indeed produce a \textit{minimum} at $(v_\phi,v_s)$.
Note that $V$ also depends on $y_t$, $g$ and $g'$ through the Coleman-Weinberg term.
We derive the values of these parameters at the $\bar{\mu}$ scale from the well-known SM one-loop RG equations.

Let us note that the outlined procedure does not take into account loop corrections to the masses of scalar particles (not to be confused with the VEVs $v_\phi$ and $v_s$ which are set to the desired values at the one-loop level).
Demanding the one-loop mass of the SM-like Higgs to be equal to \SI{125}{GeV}, would significantly complicate our already nontrivial procedure for generating parameter points.
Although our results are not very dependent on the exact values of the scalar masses, we have computed one-loop corrections to the CP-even mass-squared matrix for all generated parameter points in order to check the impact of radiative corrections.
We generally find that the mass eigenvalue associated with the SM-like Higgs remains of order \SI{100}{GeV}, indicating that loop corrections are subdominant in this case.
This is expected because the impact of new physics that is close to the electroweak scale, combined with the usual loop suppression factors should not result in too large radiative contributions to the SM-like Higgs mass.
Let us also note the following interesting  property that our loop-level analysis has shown:
In most cases, the SM-like Higgs turns out to be heavier than the second eigenstate at one loop, even when it was lighter at tree-level.
This is related to radiative contributions of the $r$ field.
To be more precise, $\lambda_{sr}$, being typically the largest quartic coupling in the model, can yield significant negative corrections which may substantially reduce the tree-level mass of the eigenstate mainly consisting of the singlet field $s$.
Note that the impact of  $r$ loops to the mass of the SM-like Higgs is much weaker simply because $\lambda_{\phi r}$ is smaller than $\lambda_{sr}$ for all parameter points that we found.
Finally, note that we have also checked that mixing imposed at tree-level  is radiatively stable.  

After obtaining the parameter points for different values of $\theta_m$ and $\theta_p$, we applied the renormalization group equations (RGE) to each set of generated quartic couplings in order to identify the parameter points for which there are no Landau poles below the Planck scale\footnote{The RGEs for the considered model may be found in \cite{Helmboldt:2016mpi}.}.
Such parameter sets will be in focus of \cref{sec:results}.

\subsection{The model at finite temperatures}
\label{subsec:finiteT}
\noindent
Our investigation of the model's phase structure will be based on the daisy-improved one-loop finite-temperature effective potential $V_\text{eff}$ (see \eg \cite{Quiros1999}), whose global minimum determines the theory's true ground state, and which can be written as
\begin{align}
	V_\text{eff}(\phi_c,s_c,T) = V_0(\phi_c,s_c) + V_\tinytext{CW}(\phi_c,s_c) + V_\tinytext{FT}(\phi_c,s_c,T) + V_\text{ring}(\phi_c,s_c,T) \fineq{.}
\end{align}
Here, \mbox{$V_0(\phi_c,s_c)=V_\text{tree}(\phi_c/\sqrt{2},s_c,0)$} with the tree-level potential from \cref{eq:Vtree}, and the Coleman-Weinberg contribution $V_\tinytext{CW}$ was already given in \cref{eq:CW}.
Employing the same field-dependent tree-level masses $m_i$ and multiplicities $n_i$ as introduced in \cref{subsec:model}, the one-loop finite-temperature contribution to the effective potential reads
\begin{align}
	V_\tinytext{FT}(\phi_c,s_c,T) = \frac{T^4}{2\pi^2} \sum_i n_i \, J_i\!\left(\frac{m_i^2(\phi_c,s_c)}{T^2}\right) \fineq{,}
\end{align}
with $J_i$ being the usual thermal functions appropriate to bosonic and fermionic loops,
\begin{align*}
	J_\tinytext{B,F}(r^2) = \int_0^\infty \!\!\dd x x^2 \log\left( 1\mp\E^{-\sqrt{x^2+r^2}} \right) \fineq{,}
\end{align*}
which can be readily approximated using Bessel functions, see \eg \cite{Kapusta2011}.
Finally, for the purpose of improving the robustness of our perturbative approach, we include the so-called ring terms into our calculation \cite{Carrington:1991hz}, namely
\begin{align}
	V_\text{ring}(\phi_c,s_c,T) = -\frac{T}{12} \sum_{i\in\text{bosons}} \!\! n_i \left( \Bigl[M_i^2(\phi_c,s_c,T)\Bigr]^{3/2} - \Bigl[m_i^2(\phi_c,s_c)\Bigr]^{3/2} \right) \fineq{.}
\end{align}
Here, each bosonic degree of freedom is supposed to have thermal mass-squared $M_i^2$, while $m_i^2$ is the corresponding zero-temperature field-dependent mass-squared from \cref{subsec:model}.
For CP-even scalars, the thermal masses $M_\pm^2$ are obtained as the eigenvalues of the matrix $m^2+\Pi$, where $m^2$ is given in \cref{eq:mass-higgs}, and $\Pi$ is the matrix of thermal self-energies with the following diagonal entries
\begin{align}
\label{eq:tm}
\begin{split}
	\Pi_\phi(T) &  = \frac{T^2}{48} \left( 24 \lambda + 2 \lambda_{\phi s} + 2 \lambda_{\phi r} + 12 y_t^2 + 9 g^2 + 3 {g^\prime}^2 \right) \fineq{,} \\
	\Pi_s(T) & = \frac{T^2}{24} \left( 6\lambda_s + 4\lambda_{\phi s} + \lambda_{sr} \right) \fineq{.}
\end{split}
\end{align}
The thermal self-energy of the $r$ field can be computed to be
\begin{align}
\Pi_r(T) = \frac{T^2}{24} \left( 6\lambda_r + 4\lambda_{\phi r} + \lambda_{sr} \right) \fineq{,} 
\end{align}
while that of the Goldstone bosons matches $\Pi_\phi(T)$ from \cref{eq:tm}.
Thermal mass-squares for $r$ and the Goldstone bosons are simply calculated as the sum of field-dependent tree-level mass-squares and the respective thermal part $\Pi$.
In the gauge sector, only longitudinal components of the SM gauge bosons contribute.
To be more precise, we have \cite{Carrington:1991hz}
\begin{align}
	\Pi_B^T(T) & = \Pi_{W_i}^T(T) = 0 \fineq{,} \quad
	\Pi_{W_i}^L(T) = \frac{11}{6} g^2 T^2 \fineq{,}\quad 
	\Pi_B^L(T) = \frac{11}{6} {g^\prime}^2 T^2 \fineq{.}
\end{align}
For the neutral gauge bosons it is convenient to work in the mass basis and identify thermal masses of $Z$ and $\gamma$ which read (see \eg \cite{Prokopec:2018tnq})
\begin{align}
	M_{Z,\gamma}^2(\phi_c,T) ={}& \frac{1}{2} \Bigl[ (g^2 + {g^\prime}^2) (\tfrac{1}{4}\phi_c^2+\tfrac{11}{6}T^2)
	\pm \sqrt{(g^2 - {g^\prime}^2)^2 (\tfrac{1}{4}\phi_c^2+\tfrac{11}{6}T^2)^2 + \tfrac{1}{4}\phi_c^4 g^2 {g^\prime}^2} \Bigr] \fineq{.}
\end{align}

In order to be phenomenologically viable, the low-temperature phase of the minimal conformal model must exhibit a vacuum that spontaneously breaks both scale-invariance and the electroweak symmetry, \mbox{$v_s \neq 0 \neq v_\phi$}, see Ref.~\cite{Helmboldt:2016mpi}.
One of the main purposes of this paper is to investigate the question of how the aforementioned vacuum may emerge from a fully symmetric ground state, \mbox{$v_s=v_\phi=0$}, in the early Universe.
The formalism to do so is well developed so we only briefly sketch it here.

We start with the general observation that the scale-symmetry-breaking phase transition in a classically conformal model is necessarily of first order, see \eg Refs.~\cite{Marzola:2017jzl,Brdar:2018num}.
This type of transition is known to proceed via the nucleation of bubbles containing the true ground state, which subsequently grow inside an expanding Universe that is still in the metastable phase.
At which temperature the phase transition completes (if at all) therefore crucially depends on the rate $\Gamma$ of bubble nucleation, on the one hand, and on the Hubble parameter $\hubble$, on the other hand.
The former quantity can be estimated as \cite{Linde1981,Hogan1984,Witten1984}
\begin{align}
	\Gamma(T) \simeq T^4 \left( \frac{\mathcal{S}_3}{2\pi T} \right)^{\!\!\nicefrac{3}{2}} \E^{-\mathcal{S}_3/T} \fineq{.}
	\label{eq:finiteT:Gamma}
\end{align}
The theory's three-dimensional Euclidean action $\mathcal{S}_3$ in the above expression is to be evaluated for the $\oN{3}$-symmetric bounce solution \mbox{$\fieldVec_\text{b}(r):=(\phi_\text{b}(r),s_\text{b}(r))^\transpose$}, which is obtained by simultaneously solving the scalar fields' coupled equations of motion,
\begin{align}
	\frac{\dd^2 \fieldVec}{\dd r^2} + \frac{2}{r} \frac{\dd \fieldVec}{\dd r} = \vec{\nabla}_{\!\mathsmaller{\Phi}} V_\text{eff} \fineq{,}
	\label{eq:finiteT:eom}
\end{align}
subject to the boundary conditions \mbox{$\fieldVec\to0$} as \mbox{$r\to\infty$} and \mbox{$\dd \fieldVec/\dd r=0$} at \mbox{$r=0$}.
In all of the above, $r$ denotes the radial coordinate of three-dimensional space.
In the context of the present paper, we use the \texttt{CosmoTransitions} code \cite{Wainwright2012} both to solve the system in \cref{eq:finiteT:eom} and to calculate the resulting action $\mathcal{S}_3[\fieldVec_b(r)]$.

As previously indicated, the second crucial quantity regarding the investigation of the phase transition is the Hubble parameter, which, in the considered scenario, can be written in terms of the Universe's radiation and vacuum energy densities $\rho_\text{rad}$ and $\rho_\text{vac}$, respectively:
\begin{align}
	\hubble^2(T) = \frac{\rho_\text{rad}(T) + \rho_\text{vac}(T)}{3 \MPl^2}
	= \frac{1}{3 \MPl^2} \left( \frac{\pi^2}{30} g_* T^4 + \Delta V(T) \right) \fineq{.}
	\label{eq:finiteT:Hubble}
\end{align}
Here, \mbox{$\MPl=\SI{2.435e18}{GeV}$} stands for the reduced Planck mass, while $g_*$ is the effective number of relativistic degrees of freedom.
The vacuum contribution to the Hubble parameter is given by the potential difference between the false ground state at \mbox{$\fieldVec=(0,0)$} and the true one at \mbox{$\fieldVec=(v_\phi(T),v_s(T))$}, that is \mbox{$\Delta V(T) := V_\text{eff}(0,0,T) - V_\text{eff}(v_\phi(T),v_s(T),T)$}.
Note that the vacuum term is typically only relevant in the case of strong supercooling, \ie if the phase transition does not complete until the Universe cools down far below the critical temperature $T_c$, at which the two aforementioned ground states are energetically degenerate.

Comparing the rate $\Gamma$ from \cref{eq:finiteT:Gamma} to the Hubble parameter $\hubble$ from \cref{eq:finiteT:Hubble} eventually allows us to estimate the temperature $T_n$ (henceforth referred to as the nucleation temperature), at which both efficient bubble nucleation and growth are possible, namely
\begin{align}
	\Gamma(T_n) \stackrel{!}{=} \hubble^4(T_n) \fineq{.}
	\label{eq:finiteT:Tn}
\end{align}
In evaluating the above condition we will ignore the factor $(\mathcal{S}_3/(2\pi T)^{3/2})$ in \cref{eq:finiteT:Gamma}, which only slowly varies with $S_3/T$ as compared to the exponential factor.
Furthermore, we will exploit the fact that whenever the vacuum contribution is relevant, it can be reliably approximated by its zero-temperature value, \ie \mbox{$\Delta V(T) \approx \Delta V(T=0)$} for all \mbox{$T\ll T_c$}.

\section{Results}
\label{sec:results}

\subsection{Relation between supercooling and RG consistency}
\label{subsec:main}
\noindent
In order to investigate the electroweak phase transition (EWPT) in the minimal scale-invariant model, we sampled \textit{phenomenologically} viable parameter points by applying the procedure outlined in \cref{subsec:magic}.
For each point we then constructed the finite-temperature effective potential presented in \cref{subsec:finiteT}.
Note that we studied both possibilities of identifying the SM-like Higgs particle with one of the eigenstates of the mass matrix from \cref{eq:mass-higgs}, \ie either with the heavier or lighter one.

\begin{figure}[t]
	\centering
	\includegraphics[scale=0.9]{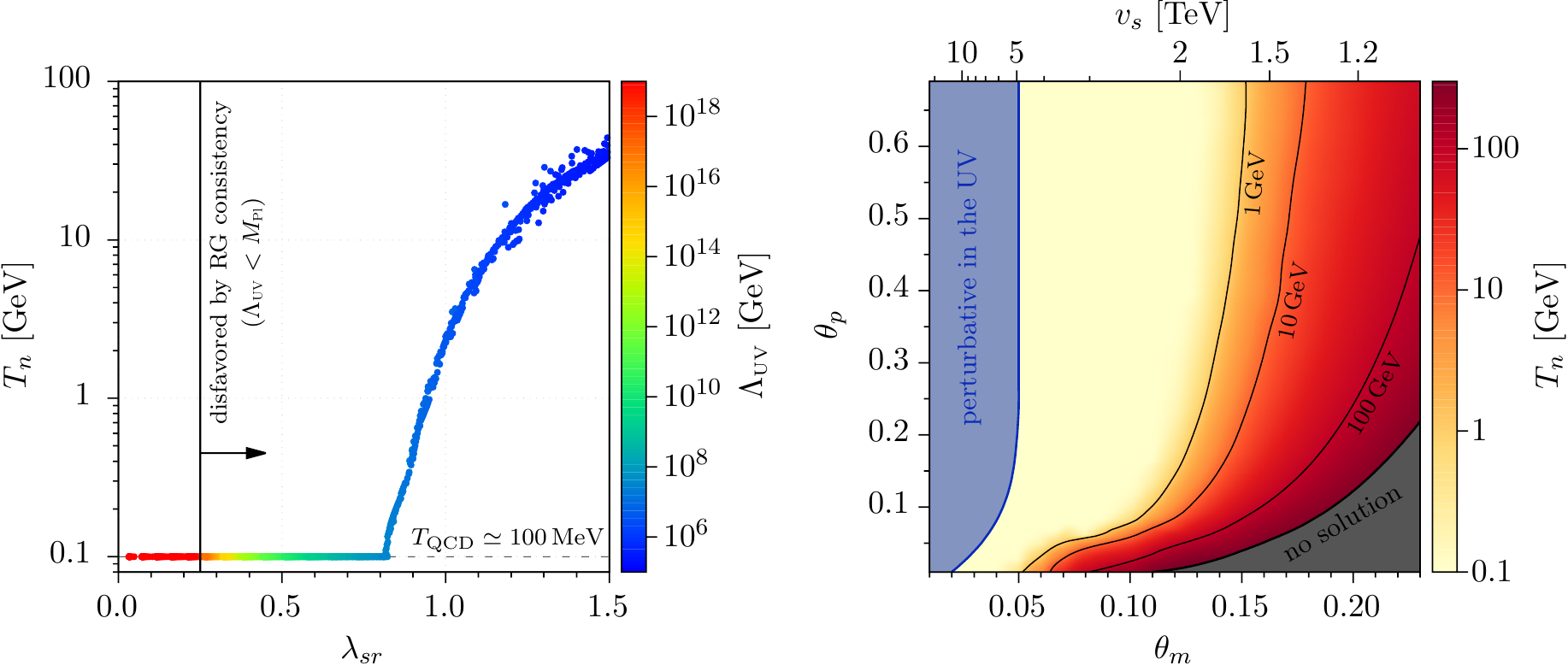}
	\caption{Temperature $T_n$ at which the electroweak phase transition (EWPT) completes in the minimal scale-invariant model.
Since the EWPT is at the latest induced by the chiral phase transition of QCD, we never show nucleation temperatures below $T_\tinytext{QCD}\simeq \SI{100}{MeV}$.
Both panels are based on the same set of parameter points, which was constructed using the method outlined in \cref{subsec:magic} and assuming that the SM-like Higgs is the lightest of the CP-even scalar eigenstates \textit{at tree-level}.
\textit{Left}: $T_n$ plotted against the portal coupling $\lambda_{sr}$. The color code indicates the lowest RG-scale $\Lambda_\tinytext{UV}$ at which a Landau pole appears.
In particular, only the red points for which \mbox{$\Lambda_\tinytext{UV} \lesssim \MPl$} are fully consistent in the sense that they can stabilize the electroweak against the Planck scale.
\textit{Right}: $T_n$ (color code and black contours) in the plane spanned by $\theta_p$ and $\theta_m$ or $v_s$, respectively.
The fully consistent region of parameters free from sub-Planckian Landau poles is shown in blue.
In the remaining part of the shown parameter space, we also present scenarios for which perturbativity is violated; in the pale region supercooling is halted only by QCD phase transition, while in the orange and red regions the EWPT completes via bubble nucleation.
While for completeness we show the full range of $\theta_p$ note that only \mbox{$\theta_p\lesssim \num{0.44}$} region is unexcluded from collider searches \cite{Farzinnia:2014xia}.}
	\label{fig:results:Tn-vs-lsr}
\end{figure}

As was already demonstrated in Ref.~\cite{Helmboldt:2016mpi}, the value of the portal coupling $\lambda_{sr}$ at the scale of radiative symmetry breaking is crucial for the successful implementation of the minimal scale-invariant model.
On the one hand, $\lambda_{sr}$ needs to be relatively large so that the second scalar singlet, $r$, is heavy enough to outweigh the top-quark and thereby stabilize the one-loop vacuum.
On the other hand, too large values lead to the appearance of Landau poles at some scale $\Lambda_\tinytext{UV}$ \textit{below} the Planck scale and thus necessarily imply the reintroduction of fine-tuning according to the arguments of Ref.~\cite{Meissner:2007xv}.
As can be seen from the left panel of \cref{fig:results:Tn-vs-lsr} where we show $\Lambda_\tinytext{UV}$ in the $\lambda_{sr}$-$T_n$ plane, it turns out that \mbox{$\lambda_{sr}\lesssim 0.3$} is required in order to avoid Landau poles below the Planck scale%
\footnote{While this is a robust statement for the considered model with extra scalars, the conclusions may  change in different non-minimal realizations of classical scale invariance.
For instance, in viable conformal models containing an extended gauge sector \cite{Prokopec:2018tnq,Marzo:2018nov} strong supercooling can be circumvented by choosing  $\gtrsim\mathcal{O}(10^{-1})$ values for the extra gauge couplings.
Unlike in the scalar case, this coupling is renormalized multiplicatively due to the protective gauge symmetry and therefore typically exhibits a more stable RG flow, so that sub-Planckian Landau poles can also be avoided for larger initial values of the gauge coupling at the TeV scale.}
(parameter points shown in red).

From the same plot, we see that the temperature $T_n$ at which the EWPT completes generally grows for increasing values of $\lambda_{sr}$.
Since the phase transition's critical temperature $T_c$ was found to always be within one order of magnitude, the value of $T_n$ gives an approximate measure for the amount of supercooling, namely smaller $T_n$ corresponds to stronger supercooling.
The left panel of \cref{fig:results:Tn-vs-lsr} now shows that we find parameter points, for which the EWPT is only moderately supercooled and can therefore complete in the usual way via bubble nucleation and percolation at temperatures up to \SI{100}{GeV} (blue points).
However, all these points feature Landau poles far below the Planck scale and must be regarded as inconsistent if one demands the hierarchy problem to be absent.

For fully consistent points with \mbox{$\lambda_{sr}\lesssim 0.3$}, the Hubble expansion parameter is larger than the bubble nucleation rate even at sub-GeV temperatures, so that \cref{eq:finiteT:Tn} cannot be satisfied.
This indicates that the Universe undergoes an extended vacuum-dominated epoch across orders of magnitude in temperature, \ie the EWPT is significantly supercooled.
However, even for very small $\lambda_{sr}$, supercooling cannot continue indefinitely.
Rather, the completion of the EWPT is induced by the chiral phase transition of QCD at temperatures of the order of the QCD scale \mbox{$T_\tinytext{QCD}\simeq\SI{100}{MeV}$} \cite{Iso:2017uuu}.
The parameter sets for which this happens are shown along the horizontal line at \mbox{$T_n=\SI{100}{MeV}$} in the left panel of \cref{fig:results:Tn-vs-lsr}.
While the aforementioned fully consistent points are drawn in red, it is furthermore interesting to note that the chiral phase transition can also halt supercooling for parameter sets that feature sub-Planckian Landau poles.

The physical picture for the EWPT triggered by QCD effects is as follows:
The QCD chiral phase transition with six massless flavors leads to the formation of chiral quark condensates, which, via Yukawa interactions with the SM Higgs field, induce terms linear in $\phi$ \cite{Witten:1980ez}.
Due to its large Yukawa coupling \mbox{$y_t\simeq1$}, the dominant contribution comes from the top quark condensate.
The term linear in $\phi$ subsequently induces a finite vacuum expectation value for that field, namely \mbox{$v_\phi^\tinytext{QCD} \simeq \mathcal{O}(\SI{100}{MeV})$}, which spontaneously breaks electroweak symmetry.
Such a VEV then generates a mass term for the $s$ field, \smash{$m_s^2=(\lambda_{\phi s}/2)[v_\phi^\tinytext{QCD}]^2$}.
Since $\lambda_{\phi s}$ is negative for all of our parameter points, this mass counteracts the thermal self-energy $\Pi_s$ (expression given in \cref{eq:tm}).
As the temperature drops, the thermal contribution ceases and at the point when the two contributions are equal (matching absolute values, signs are still different), the $s$ field starts rolling down the potential toward the true minimum, provided that the field did not already tunnel beforehand in a first-order phase transition.
Numerically, we have found that this occurs between \mbox{$T_\text{end}=\SI{8}{MeV}$} and \SI{33}{MeV} for \mbox{$v_\phi^\tinytext{QCD}=\SI{100}{MeV}$}.
Clearly, such temperatures are still above those at which Big Bang Nucleosynthesis (BBN) results can constrain new physics, making the outlined cosmological scenario viable.
Note that $T_\text{end}$ is proportional to $v_\phi^\tinytext{QCD}$, so that BBN limits could only play a role if the QCD scale was an order of magnitude smaller than expected, which is essentially inconsistent with QCD lattice results \cite{Montvay:1994cy,Gockeler:2005rv}.
Once the $s$ field has settled at its minimum, the known electroweak scale \mbox{$v_\phi=\SI{246}{GeV}$} emerges via the portal coupling $\lambda_{\phi s}$.

In the right panel of \cref{fig:results:Tn-vs-lsr} we show the nucleation temperature in the $\theta_p$-$\theta_m$ plane.
The part of parameter space, in which the model features perturbativity up to the Planck scale and thus may accommodate a solution to the hierarchy problem, is shown in blue.
As already elaborated, it is QCD effects that induce electroweak and scale symmetry breaking in this region.
The corresponding angles $\theta_m$ are rather small (equivalent to large values of $v_s$).
This is expected as the mass of the $r$ field grows with $\lambda_{sr} v_s^2$, so that for larger values of $v_s$ smaller couplings suffice to stabilize the one-loop vacuum.
Hence, the blue region corresponds to the previously discussed \mbox{$\lambda_{sr}\lesssim 0.3$} window.
In the same plot one can also observe a relatively large part of parameter space in which Landau poles do appear below the Planck scale, but the chiral phase transition is still responsible for the generation of the electroweak scale (pale color).
As $v_s$ further decreases, the region in which the EWPT completes via bubble nucleation at temperatures above the QCD scale is reached.
However, as already argued, the model cannot stabilize the electroweak against the Planck scale in this part of parameter space.
Note also the existence of a region in which we found no solutions using our approach of \cref{subsec:magic}.

\subsection{Dark Matter and Gravitational Waves}
\label{subsec:DM-GW}
\noindent
The DM candidate in the model is the scalar gauge singlet $r$ which is stable due to the imposed $\mathbb{Z}_2$ symmetry.
The authors of Ref.~\cite{Hambye:2018qjv} have proposed that a significant fraction of DM may come from a \enquote{supercool} component.
This is essentially a thermal abundance of the massless $r$ field that gets diluted during the supercooling phase when the Universe expands exponentially.
Eventually, the period of supercooling ends and the phase transition completes, so that $r$ becomes massive with the dark matter yield of $\left[45/(2\pi^4 g_*)\right] (T_\text{end}/T_\text{infl})^3$.
Here, $T_\text{infl}$ is the temperature below which the Universe starts being vacuum dominated during supercooling.
In our model it equals the temperature $T_\tinytext{RH}$ to which the thermal plasma is reheated after the phase transition has completed.
For all viable benchmark points that do not feature Landau poles below the Planck scale, we have found that the \enquote{supercool} DM abundance cannot account for the total DM abundance.
Furthermore and more importantly, we also found that $T_\tinytext{RH}$ is always larger than the freeze-out temperature (this statement is independent of the value of $v_\phi^\tinytext{QCD}$).
This essentially implies that the \enquote{supercool} abundance in the considered scenario does not leave any imprint and is effectively washed out since after reheating, $r$ undergoes standard freeze-out.
While we have identified several benchmark points that yield a consistent DM abundance from freeze-out, we do not perform a detailed DM study here, but rather refer the interested reader to the vast literature on Higgs portal DM models (see \eg review \cite{Arcadi:2019lka} and references therein).

Lastly, let us comment on gravitational wave signatures that may originate in this model from potentially strong first-order cosmic phase transitions \cite{Witten1984,Hogan1984,Hogan1986b,Turner1990a,Kamionkowski1993}.
Since the electroweak phase transition for all fully consistent parameter points is delayed down to sub-GeV temperatures, the QCD phase transition occurs while quarks are still massless.
Such a transition is known to be of first order \cite{Pisarski1984a} and was shown to produce a stochastic gravitational wave background in the range of proposed near-future detectors, provided the chiral phase transition does not proceed too quickly \cite{Schwaller:2015tja,Iso:2017uuu}.
However, recent explicit calculations indicate that the transition does complete very fast, so that the associated gravitational waves signal may be too weak to be observable \cite{Helmboldt:2019pan}.
Let us also note that there could potentially be another first-order phase transition associated with the spontaneous breakdown of scale symmetry and following the QCD phase transition.
As already briefly discussed in \cref{subsec:main}, after the generation of a finite $v_\phi^\tinytext{QCD}$, the $s$ field will either roll down the potential or undergo a first-order phase transition.
While we do not study which of the two scenarios occurs for our parameter points, note that the latter option is not expected to produce an observable gravitational wave signature \cite{Baldes:2018emh}, implying that both cosmological scenarios are phenomenologically equivalent.

\section{Summary and Conclusions}
\label{sec:summary}
\noindent
In the absence of any new findings at the LHC, the gauge hierarchy problem remains one of the greatest challenges in high-energy physics.
Among various proposals for its solution, classically scale-invariant theories belong to the most minimal options, typically requiring only rather simple extensions of the Standard Model (SM) particle content.
In this paper we explored the electroweak phase transition (EWPT) in the scale-invariant model in which the SM is supplemented with two extra scalar gauge singlets.
This model was previously shown to offer the minimal phenomenologically consistent framework. 
Let us point out that the analysis techniques employed to investigate radiative symmetry breaking in the literature chiefly boil down to the Gildener-Weinberg approach relying on the existence of exact flat directions in the tree-level potential.
In this work, however, we took a complementary and more general approach in which both tree-level and radiative terms in the potential play a role in the generation of the electroweak scale.

We argued that consistently avoiding the hierarchy problem in the model requires the absence of sub-Planckian Landau poles.
As a consequence, we found that the portal couplings of the viable parameter points must necessarily be smaller than $\mathcal{O}(10^{-1})$.
This has a rather significant imprint on the physics of the early Universe.
In particular, we found that with such small couplings, the nucleation rate of critical bubbles containing the true electroweak vacuum cannot compete with the Hubble expansion even at relatively low temperatures.
The EWPT can therefore not complete conventionally via bubble nucleation.
Instead, the chiral phase transition of QCD plays a crucial role in inducing the EWPT and generating the electroweak scale. 
Before reaching the QCD phase transition temperature, the Universe experiences an epoch of vacuum-dominated expansion, in which it is still in the symmetric phase.
In other words, the EWPT is strongly supercooled.
The amount of supercooling decreases if larger portal couplings are considered.
However, the model is then no longer perturbative all the way up to the Planck scale and can thus not avoid the gauge hierarchy problem.

 We conjectured that the described relation between renormalization group
consistency and strongly supercooled scale-generating phase transitions
is generally true in purely scalar classically scale-invariant
extensions of the SM.
In contrast, strong supercooling may be prevented in scale-invariant
gauge extensions of the SM by choosing large enough gauge couplings.

Let us stress that even though supercooling is usually associated with a rather strong gravitational wave signal, particularly in the context of scale-invariant models, we concluded that there would be no testable stochastic gravitational wave background produced in association with the aforementioned cosmology in the considered minimal scale-invariant model.

\bibliographystyle{JHEP}
\bibliography{refs}

\end{document}